\documentclass[review]{elsarticle}
\usepackage[hidelinks]{hyperref}

\biboptions{sort&compress}

\usepackage{amsfonts}
\usepackage{graphicx}
\usepackage{amsmath}
\usepackage{subcaption}
\usepackage{nomencl}
\usepackage{etoolbox}
\usepackage{booktabs}
\usepackage[linesnumbered,ruled,vlined]{algorithm2e}
\SetKwFor{For}{for }{ do }{end}
\SetKwInput{KwInput}{Input}                
\SetKwInput{KwOutput}{Output}              

\makenomenclature

\newcommand{\reals}{\mathbb{R}}

\begin{document}

\begin{frontmatter}
\title{Robust co-design framework for buildings operated by predictive control \tnoteref{t1}}
\tnotetext[t1]{This work was supported by the Engineering and Physical Sciences Research Council (EPSRC) [grant EP/V012053/1];}

\author[First]{P. Falugi \corref{mycorrespondingauthor}} \cortext[mycorrespondingauthor]{Corresponding author}
\ead{p.falugi@uel.ac.uk}
\author[Second]{E. O'Dwyer}
\author[Third]{M. A. Zagorowska} 
\author[Fourth,Fifth]{E. C. Kerrigan }
\author[Sixth]{Y. Nie}
\author[Fifth]{G. Strbac}
\author[Second]{N. Shah}

\address[First]{Engineering and Construction Department, University of East London, London, UK}
\address[Second]{Department of Chemical Engineering, Imperial College London, London, UK}
\address[Third]{Mechanical Engineering Department, Delft University of Technology, Delft, The Netherlands}
\address[Fourth]{Department of Aeronautics, Imperial College London, London, UK}
\address[Fifth]{Department of Electrical and Electronic Engineering, Imperial College London, London, UK}
\address[Sixth]{Department of Automatic Control and Systems Engineering, University of Sheffield, UK}

\begin{abstract}
Cost-effective decarbonisation of the built environment is a stepping stone to achieving net-zero carbon emissions since buildings are globally responsible for more than a quarter of global energy-related CO$_2$ emissions. 

Improving energy utilization and decreasing costs naturally requires considering multiple domain-specific performance criteria.
The resulting problem is often computationally infeasible.

The paper proposes an approach based on decomposition and selection of significant operating conditions to achieve a formulation with reduced computational complexity. 
We present a robust framework to optimise the physical design, the controller, and the operation of residential buildings in an integrated fashion, considering external weather conditions and time-varying electricity prices.  The framework explicitly includes operational constraints and increases the utilization of the energy generated by intermittent resources.

A case study illustrates the potential of co-design in enhancing the reliability, flexibility and self-sufficiency of a system operating under different conditions. Specifically, numerical results demonstrate reductions in costs up to $30$\% compared to a deterministic formulation. Furthermore, the proposed approach achieves a computational time reduction at least $10$ times lower compared to the original problem with a deterioration in the performance of only 0.6\%.
\end{abstract}

\begin{keyword}
\texttt{elsarticle.cls}\sep \LaTeX\sep Elsevier \sep template
\MSC[2010] 00-01\sep  99-00
\end{keyword}

\end{frontmatter}
\section{Introduction}

Cost-effective achievement of net-zero carbon emissions is one of the most critical challenges due to the exponential growth of energy demand and multiple uncertainties affecting the production and usage of energy. 
Active participation of buildings in the whole energy system to achieve emission targets is extremely valuable since commercial and residential buildings are responsible for about 38\% of the global CO$_2$ emissions \cite{Build_emiss_20}.  
Active support of buildings for the whole energy system necessitates adopting renewable resources, various storage technologies and smart devices. It also requires the participation of consumers in dynamic electricity markets characterised by time-varying prices. 
The design of flexible and cost-efficient buildings requires optimising the technologies' size considering the dweller needs and how the system operates under different possible conditions.
Frequently, the effect of how the system is operated is neglected in the design process, usually because including the optimal closed-loop operation of the system in the sizing problem and considering the uncertainty in the operating conditions leads to a computationally challenging problem. 
Nevertheless, neglecting operational uncertainty and the behaviour of closed-loop systems may lead to a suboptimal design.

Motivated by the requirement for the buildings to increase their energy efficiency at affordable costs, we propose a framework for optimally designing the system and the controller parameters considering uncertain operating conditions. 

Given the desired high performance, the presence of constraints and the absence of significant nonlinearity, Model Predictive Control (MPC) is the natural control choice for operating buildings.
Furthermore,  the availability of near-term weather forecasts and the prospective participation of consumers in dynamic electricity markets characterised by time-varying prices makes Economic Model Predictive Control (EMPC) \cite{Angelietal2016,Risbecketal2020} the natural choice for a dweller to improve performance while optimising the electricity cost. 

The co-design approach, proposed in this work, includes tuning the controller parameters, such as prediction horizon, sampling time and discretization step, looking at closed-loop performances.

\subsection{Existing literature}

\subsubsection{Controller tuning}

Traditionally, the success of the tuning process relied on experience.
Recently, tuning approaches based on the optimisation of closed-loop performance have attracted the attention of the scientific community. In \cite{Bachtiaretal2017}, an offline calibration of a model-predictive integrated missile controller was performed, seeking a compromise between closed-loop performance and computational capacity in a deterministic setting. The method tunes the MPC sampling time and the prediction horizon, evaluating the multi-objective function at different points to find a trade-off between performances and computational requirements. The solution method relies on the continuity property of the tuning objective functions and extends the approach developed in \cite{Bachtiaretal2016} for the linear case to the nonlinear case. 

Continuity properties of the open-loop and closed-loop value functions with respect to the sampling and horizon times have been demonstrated for linearly constrained quadratic optimisation problems in  \cite{Bachtiaret2al2016}. Bayesian optimization (BO) using surrogate functions has been applied to MPC calibration in \cite{Forgioneetal2020}.
In \cite{Lucchinietal2020}, tuning is achieved using the global optimization based on inverse distance weighting and radial basis function surrogates algorithm proposed in \cite{Bemporad2020}.
Preference-based optimisation approaches have been applied to MPC automatic calibration in \cite{zhu2021cglisp} when the objective and the constraints are unknown. Preference-based decision processes have the advantage of not requiring the definition of a merit index and the parameter choice can be performed using qualitative criteria. The merit index is replaced by a human operator providing preferences, and the computational burden of evaluating closed-loop performance to optimality is reduced.

A semi-automated calibration approach for multi-objective problems is proposed in \cite{zhu2021preferencebased} based on learning a surrogate of the latent objective function and making a decision on preferences \cite{Bemporadetal2021}. 
Recently a BO approach for the tuning of time-invariant MPC applied to heating, ventilation, and air conditioning (HVAC) plants is proposed in \cite{lu2021mpc}. The approach in \cite{lu2021mpc} aims to quickly and optimally tune the controller by exploring the space of the tuning parameters. 
In \cite{Ericetal_2019}, an automatic tuning of the MPC is proposed in a co-design framework to achieve an optimal trade-off between performances and computational resources in nominal conditions. Their approach automates the MPC and hardware co-design using Bi-objective Mesh Adaptive Direct Search algorithm (BiMADS) to handle discrete variables.

\subsubsection{Co-design}

Recently, \cite{Sanz_19} recognized the importance of adopting a control co-design framework to push the performance of a system to its achievable limits.  The control co-design framework considers multidisciplinary subsystem interactions in a unified manner, enabling the opportunity to improve performance substantially. In the 1980s, the co-design idea was integrated with optimization schemes to identify control and parameters \cite{Rao_1988}.
Frameworks co-optimising closed-loop software implementation and hardware performance appear in \cite{Suardietal_2013,Kircheretal2019} where the co-design consists of a multi-objective formulation since the decision variables span different time scales and belong to different areas.  A survey on control co-design applications is provided in \cite{Process_Diangelakis2017}. 
The optimal selection of the equipment considering the system operation is critical to increase the energy efficiency of large heating systems as pointed out in \cite{Henzeetal2008,Powelletal2013}. 

\subsection{Contributions} 
This paper considers a robust co-design framework for residential buildings operated by a time-varying economic MPC where continuity assumptions are not satisfied. In the building sector, the optimal equipment combination, independently of the pricing mechanism, significantly increases the achievable energy efficiency, as highlighted in 
\cite{Risbeck2etal2020} for the online control of large-scale HVAC in commercial buildings.
Performance strongly depends on uncertain operating conditions, the closed-loop MPC controller, and the choice of controller parameters.  
The required prediction horizon length depends on the timescales of the system and periodic components.
Consequently, the present contribution proposes a framework for automating the technology choice and the tuning of the EMPC operating in closed-loop, considering the effect of uncertainty.
The current work mainly accounts for the effect of large uncertainty described by different possible operating conditions, external weather conditions and dynamic electricity prices. 
However, the framework can handle other uncertainty descriptions at the expense of increased computational complexity.

\subsubsection{Novelty} Given the complexity of the problem, further exacerbated by uncertainty, we propose learning techniques to select relevant data for the  solution of the problem.
A naive choice of subsets of data in most cases results in a solution far from optimal \cite{Hoffmannetal2020,Hilbersetal2020}.
Many selection methods and aggregation techniques select and aggregate typical subsamples without considering what might be relevant to the problem under consideration, and they easily neglect extreme events. 
Consequently, the obtained designed choices can perform poorly in such operating conditions.     
Recent importance subsample techniques \cite{HILBERS2019113114,Hilbersetal2020} identify extreme events that are significant for the problem of interest, achieving a design choice capable of good performance on a more extensive set of operating conditions at a reduced computational burden. 
In \cite{HILBERS2019113114,Hilbersetal2020} the importance subsamples are chosen looking at their effect on the problem output. 
In the present framework, we extend the idea of the importance of subsample selection. Since time-varying electricity prices induce a large cost variability, we determine the importance of a subsample based on optimal costs  and sizing choice. The introduction of scaling parameters enables to balance preference for the clustering choice.

\subsubsection{Methodology} The robust co-design framework under consideration in this work has to deal with discontinuities due to the discrete nature of the sizing parameters and the time-varying piece-wise constant prices in EMPC cost. Since the prices are piece-wise constant, the variation of the optimal cost with respect to time is discontinuous. Other discontinuities can arise depending on the adopted robust MPC  formulation, such as a worst-case cost. Discontinuous black-box optimization problems involving a  moderate number of granular variables can be effectively solved using derivative-free optimization algorithms. 
Consequently, the focus of the present contribution is on robust co-design formulations, in which the Mesh Adaptive Direct Search  (MADS) algorithm takes care of the ``system+controller" parameter design, but a derivative-based solver is used to solve the MPC problem in a receding horizon fashion.
The advantage of using MADS is the possibility of directly handling systems with discontinuities.

Given the challenging nature of the problem, we propose a reformulation
through a novel importance subsampling selection approach and decomposition techniques enabling parallel implementations. The proposed approach aims to achieve an optimal trade-off between computational burden and closed-loop system performance. The proposed framework and solution methodology to solve closed-loop design for time-varying EMPC formulations with uncertainties and derivative-free solvers has not yet been proposed in the
literature. A case study shows the possible benefits of the proposed robust co-design framework considering the sizing of photovoltaic panels (PV) and batteries in residential buildings, and the uncertainty in the weather conditions and electricity prices.

The rest of the paper is structured as follows. Section \ref{sec:ProblemFormulation} introduces the problem formulation and the numerical challenges due to uncertainty and possible discontinuities. Section \ref{sec:ProblemDecomposition} presents the decomposition techniques adopted in this paper to improve the numerical properties of the algorithm and describes the proposed co-design framework. Section \ref{sec:CaseStudy} presents numerical results of the proposed framework applied to a residential building case study. The paper ends with conclusions in Section \ref{sec:Conclusions}.

\section{Problem Formulation}
\label{sec:ProblemFormulation}

\subsection{System dynamics}
The system to be designed and controlled is described by
\begin{equation}\label{eq:1}
    \dot{x}(t)=f(x(t),u(t),w(t),p,t)
\end{equation}
where $f(\cdot)$ is continuous,  $x(t) \in\reals^{n_x}$ is the system state, $u(t)\in \reals^{n_u}$ the control input, $p \in \reals^{n_p}$ a vector of constant design parameters and $w(t) \in \reals^{n_w}$ a vector of uncertain time-varying exogenous inputs belonging to a time-varying uncertainty set $\mathbb{W}(t)$. The exogenous inputs $w(t)$ define the operating conditions of the system.
The aim is to design a system and a controller to achieve high performance under plausible operating conditions described by $w(\cdot)\in \mathcal{W} : =\{\textrm{measurable bounded functions } w :[0, \infty ) \rightarrow \reals^{n_w}  \;: \;   w(t) \in  \mathbb{W}(t) \}$.
The uncertainty on the short time scale in real-time forecasting is assumed to be implicitly counteracted by the MPC controller operated in closed-loop. We are primarily interested in large uncertainties captured by scenarios characterizing significant operational differences. Given the sizing parameter $p$ and the prediction horizon $t_f$ the closed-loop controller at time $t$ minimizes the performance cost

\begin{equation}\label{cost_contr}
\displaystyle{ \int_{t}^{t+t_f} \ell(x(\tau),u(\tau),w(\tau),p,\tau)\mathrm{d}\tau}	
\end{equation}
while satisfying  the constraints 
\begin{align}
	 \dot{x}(\tau)=f(x(\tau),u(\tau), w(\tau),p,\tau) &, \quad \forall \tau \in\mathcal{T} \ \text{a.e.}, \label{eqn:dyns}\\
	g(\dot{x}(\tau),x(\tau),\dot{u}(\tau),u(\tau), w(\tau),p,\tau) \leq 0,\  & \quad \forall \tau \in\mathcal{T} \ \text{a.e.}, \label{eqn:ineqs}\\
	c(\dot{x}(\tau),x(\tau),\dot{u}(\tau),u(\tau),p, \tau) = 0,\  & \quad  \forall \tau \in \mathbb{T},  \label{eqn:wp_const}\\
	\psi(x(t),x(t+t_f),p,t,t_f)\leq 0, & \label{eqn:terminal_ineq}
\end{align}
where  `a.e.' stands for `almost everywhere' in the Lebesgue sense on the interval $\mathcal{T}:=[t,t+t_f]\subset \mathbb{R}$, where $t$ and~$t_f$ denote, respectively, the initial time and the prediction horizon. The set $\mathbb{T} \subset \mathcal{T}$ is a finite subset of $\mathcal{T}$. 
The inequality constraints \eqref{eqn:ineqs}, where $g:\mathbb{R}^{n_x} \times \mathbb{R}^{n_x} \times \mathbb{R}^{n_u} \times \mathbb{R}^{n_u} \times \mathbb{R}^{n_p} \times \mathbb{R} \to \mathbb{R}^{n_g}$ and $n_g$ indicates the number of constraints, describe general constraints on the variables.
General boundary equality and inequality constraints can be imposed through  \eqref{eqn:terminal_ineq}, where
$\psi:\mathbb{R}^{n_x} \times \mathbb{R}^{n_x} \times \mathbb{R}^{n_p} \times \mathbb{R} \times \mathbb{R} \to \mathbb{R}^{n_I}$ and $n_I$ is the number of terminal constraints.

It is assumed that the states and inputs lie in compact sets $\mathbb{X}$ and $\mathbb{U}$, respectively, which are included in (\ref{eqn:ineqs}).
The optimal closed-loop policy at time $t$ is obtained by minimizing
(\ref{cost_contr}) subject to the constraints (\ref{eqn:dyns})--(\ref{eqn:terminal_ineq}). Such a closed-loop policy, and consequently its performance, is determined by the system design parameters $p$, the realization of the uncertainty $w(\cdot)$ in the interval $[t, ~t+t_f]$ and the implemented algorithm computing the optimal controller, which necessitates the selection of several parameters such as prediction horizon, discretization step and sampling time. The choice of these parameters in a systematic way is the objective of the algorithm proposed in this work.

\subsection{Risk measures} The system is required to provide good performances under a large variety of operating conditions that are unknown at the design stage. 
The cost (\ref{cost_contr}) changes for different realizations of the uncertainty, and consequently, a measure of risk providing a surrogate for the overall cost is adopted as discussed in \cite{Rockafeller_2014}.
In particular, the measure of risk  assigning a single value to an uncertain variable $Z$
is a functional $\mathcal{R} : Z \rightarrow \; (-\infty,\infty) $, which is required to be a coherent measure of risk to be a good risk quantifier \cite{Rockafeller_2014}. Examples of coherent risk measures are the expectation $E[Z]$ and  the worst case realization $\max(Z)$.
The problem under consideration, optimizing the performance of the whole system, naturally requires multiple objectives involving performance criteria depending on the specific domain area of the different decision variables.
Without loss of generality, this paper focuses on three objectives (Subsection \ref{multi_obj}) of particular interest in the building sector.

\subsection{Numerical solution} An algorithmic implementation of the optimization problem (\ref{cost_contr})--(\ref{eqn:terminal_ineq}), given the information on the system state and its environment, computes a control law at time instants $t=t_i$ for $i=0,1,\ldots ~ $ and imposes the conditions (\ref{eqn:dyns})--(\ref{eqn:terminal_ineq}) on a discrete set $\mathcal{T}^d :=\{ \tau_0, \tau_1,\ldots, \tau_N  \}$ of time instances satisfying  $\tau_0=t_i < \tau_1 <\ldots < \tau_N=t_i+t_f $. Denote $T^{(s)}:=t_{i+1}-t_i$ as the sampling time for all $i$ and  $T_{k}^{(d)}:=\tau_{k+1}-\tau_k$ the discretization steps where $\tau_k \in  \mathcal{T}^d$, $k=0,1,\ldots, N$. For the sake of simplicity, we assume that there exists a positive integer $n_d$ such that $T^{(s)}=n_d T_{0}^{(d)}$.  
The parameter $T^{(s)}$ describes how often new measurements are retrieved and an optimization problem is solved to compute the control law.  
The finite parametrization of the problem involves a finite parametrization of all the trajectories that we highlight, adding the tilde symbol $\tilde{~}$ to their definition.
The control algorithm, given the  state $x_i$ at time $t_i$ and the sequence $\mathbf{\tilde{w}}_{[\tau_0,\tau_N]}=( \tilde{w}(\tau_0),\tilde{w}(\tau_1),  \ldots, \tilde{w}(\tau_N)) $ over the horizon $N$, minimizes a discretized formulation of the problem (\ref{cost_contr}) -- (\ref{eqn:terminal_ineq}) with a cost 
\begin{equation}
	J^{(D)}(\mathbf{\tilde{u}},\mathbf{\tilde{w}}_{[\tau_0,\tau_N]},t,t_f,p,x_i)=\sum_{k=0}^N \ell^{D}(\tilde{x}(\tau_k),\tilde{u}(\tau_k),\tilde{w}(\tau_k),p,t)
\end{equation}
that depends on the discretization method \cite{betts2010practical}. 
We denote as 
\begin{equation}\label{eq:opt_u}
\begin{array}{l}
    \mathbf{\tilde{u}^*}(x_i,t_i;\mathbf{\tilde{w}}_{[\tau_0,\tau_N]},p,p_c)=\\
    \; ( \tilde{u}^*(\tau_0;\mathbf{\tilde{w}}_{[\tau_0,\tau_N]},x_i,p,p_c),\tilde{u}^*(\tau_1;\mathbf{\tilde{w}}_{[\tau_0,\tau_N]},x_i,p,p_c),\ldots,\tilde{u}^*(\tau_{N};\mathbf{\tilde{w}}_{[\tau_0,\tau_N]},x_i,p,p_c))
  \end{array}    
\end{equation}    
and
 \begin{equation}\label{eq:opt_x}
\begin{array}{l}   
 \mathbf{\tilde{x}^*}(x_i,t_i;\mathbf{\tilde{w}}_{[\tau_0,\tau_N]},p,p_c)=\\
  \; ( \tilde{x}^*(\tau_0;\mathbf{\tilde{w}}_{[\tau_0,\tau_N]},x_i,p,p_c),\tilde{x}^*(\tau_1;\mathbf{\tilde{w}}_{[\tau_0,\tau_N]},x_i,p,p_c),\ldots,\tilde{x}^*(\tau_{N};\mathbf{\tilde{w}}_{[\tau_0,\tau_N]},x_i,p,p_c)) 
  \end{array}    
\end{equation} 
the optimal input and state sequences, respectively, returned by the control algorithm. The optimal sequences (\ref{eq:opt_u}) and (\ref{eq:opt_x}) depend on the vector $p_c$ of the controller parameters of interest, such as the prediction horizon, the sampling time and the discretization step.

The control law  $\kappa_N(t,x_i,t_i;\mathbf{\tilde{w}}_{[\tau_0,\tau_N]},p,p_c)$,
applied to the system, is
determined by the first elements of the sequence   $\mathbf{\tilde{u}^*}(x_i,t_i;\mathbf{\tilde{w}}_{[\tau_0,\tau_N]},p,p_c)$  in the time interval  
$t \in [t_i,t_i+ T^{(s)}]$ as follows
\begin{equation}\label{feedback}
	\begin{array}{l}
 \kappa_N(t,x_i,t_i;\mathbf{\tilde{w}}_{[\tau_0,\tau_N]},p,p_c):= \\ \quad \phi(t,[\tilde{u}^*(\tau_0;\mathbf{\tilde{w}}_{[\tau_0,\tau_N]},x_i,p,p_c),
\ldots,\tilde{u}^*(\tau_{\overline{k}};\mathbf{\tilde{w}}_{[\tau_0,\tau_N]},x_i,p,p_c)]) 
\end{array}
\end{equation}
where $\overline{k}$ satisfies $\tau_{\overline{k}} = t_i+T^{(s)}$ and the interpolation function $\phi(\cdot)$ is determined by the discretization method that was used. 
The achieved closed-loop cost $V_{cl}(\cdot)$ in the interval $[t_i, t_i+T^{(s)}]$ is
\begin{equation}\label{discr_cost}
\begin{array}{l}
    V_{cl}(x_i,t_i,\mathbf{w}_{[t_i,t_{i+N}] } ,p,p_c)=\\
	\quad \displaystyle{ \int_{t_i}^{ t_i+T^{(s)}} \ell(x(\tau),\kappa_N(\tau,x_i,t_i; \mathbf{\tilde{w}}_{[t_i,t_{i+N}]},p,p_c)  ,w(\tau),p)\mathrm{d}\tau}
\end{array}	
\end{equation}  
where $x(t)$ refers to the closed-loop trajectory obtained by applying the control law (\ref{feedback}) and  $\mathbf{w}_{[a,b]}$ is the vector of exogenous signals in the interval $[a,b]$.  

\subsection{Multi-objective co-design problem}
\subsubsection{Multiple objectives}\label{multi_obj}
An essential performance criterion in designing and operating a building is its economic cost, which consists of investment and operating costs given by time-varying electricity prices. The first objective of the co-design problem is, therefore, the minimization of the system present value cost \cite[Appendix A]{conejoetal2016} given by 
\begin{equation}\label{economic} 
J^{(1)}(p,p_c):=\mathcal{R}\left(\sum_{i=0}^{N_y}V_{cl}(x_i,t_i,\mathbf{w}_{[t_i,t_{i+N}]},p,p_c) +V_I(p)\right)
\end{equation}
with a chosen risk measure $\mathcal{R}$, where $V_I(p)$ takes into account investment costs and $N_y$ is the number of samples required to cover a whole year. The initial condition $x_i$ can also be considered as an uncertain parameter.
Since the performance accounts for a time-varying economic criterion the choice of the controller parameters requires the inclusion of additional objectives considering the performance of the closed-loop system.

Even though extensive research has been performed on EMPC, little has been said about closed-loop performance for the time-varying case \cite{Angelietal2016,Gruneetal2017}. 
Without any terminal constraint, economic MPC can exhibit pathological closed-loop behaviour, often arising from a mismatch between the predicted open-loop solutions and the resulting closed-loop evolution of the system. 
The authors in \cite{Risbecketal2020} provide a general framework for time-varying EMPC problems.
Closed-loop performance and stability properties are discussed under different assumptions.
Notably, the paper provides closed-loop performance bounds for an arbitrary feasible trajectory for the general case where the continuity of the cost function might not be satisfied. 
The result in \cite{Risbecketal2020} is particularly useful since it rules out the occurrence of such behaviour without enforcing convergence to the specified reference trajectory.

In general, it is not  straightforward to identify feasible trajectories and, consequently a reference performance
$V_{cl}(x_j^r,t_j,\mathbf{w}_{[t_j,t_{j+N}]},p,p_c^r)$
for $j=0, \ldots , N_y^r$ for time-varying formulations. The superscript $r$  denotes a reference term.
For this reason, the reference performance is determined by physically meaningful controller parameters from which we expect to achieve the best closed-loop performance defined by (\ref{cost_contr}).
With such a parameter choice, the controller would not be able to operate in real-time. Indeed the sensitivity analysis of the optimal cost with respect to these parameters can also indicate if their choice is adequate. 
Note that a reference is deemed suitable if the associated performance is not too sensitive to small parameter changes, especially including, as an additional design parameter, a degree of tightening in the terminal constraints to enforce a robustness margin at the operational level.
Let 
\begin{equation}
 \begin{array}{l}
  \mathbf{V_M}(x_0,t_0,\mathbf{w}_{[t_0,t_{M+N-1}]},p,p_c):=\\ \quad [V_{cl}(x_0,t_0,\mathbf{w}_{[t_0,t_{N}]},p,p_c),
  \ldots,V_{cl}(x_{M-1},t_{M-1},\mathbf{w}_{[t_{M-1},t_{M+N-1}]},p,p_c)]'
 \end{array}   
\end{equation}
where $'$ indicates the transpose of a vector.
The cost accounting for the controller closed-loop performance is defined as    
\begin{equation}\label{cl_loop_performance} 
\begin{array}{l}
J^{(2)}(p,p_c):=\\
\quad \mathcal{R}(D(\mathbf{V_{N_y}}(x_0,t_0,\mathbf{w}_{[t_0,t_{N_y+N-1}]},p,p_c),\mathbf{V_{N_y}}(x_0,t_0,\mathbf{w}_{[t_0,t_{N_y+N-1}]},p,p_c^r)))
\end{array}
\end{equation}
where  $D(\cdot)$ denotes the norm of choice. 
Other costs (see \cite{yanboetal2022})  would be relevant to perform the auto-tuning of the controller, but for simplicity, we consider that the discretization time can only assume values for which the approximating error of the system dynamics is acceptable. 
However,  a sensitive optimal cost reveals inadequate control parameter choices.

Another important tuning criterion is the computational resources required by the controller.
The MPC tuning requires a trade-off between closed-loop performance improvements and the additional computational effort needed to provide such performance improvements.
Long prediction horizons and small sampling and discretization steps might not be necessary to achieve good performance and such parameters determine the complexity of the problem to be solved.  
Consequently, the third cost accounts for the preference for short horizons and large sampling and discretization steps
\begin{equation}\label{MPC_param} 
	J^{(3)}(p,p_c):=\mathcal{R}(Q(p_c;\mathbf{w}_{[t_0,t_{N_y}]}))
\end{equation}
where $Q(\cdot)$ indicates a cost of choice depending on the controller parameters. For example, a possible choice consists of the computational time required to compute the objective
$ \mathbf{V_{N_y}}(x_0,t_0,\mathbf{w}_{[t_0,t_{N_y+N-1}]},p,p_c)$ that highlights  the potential dependence  of the cost $Q(\cdot)$ on the specific realization of $\mathbf{w}_{[t_0,t_{N_y}]}$.

\subsubsection{Co-design problem} 

\begin{figure}[!tb]
	\centering
	\includegraphics[width=0.5\textwidth]{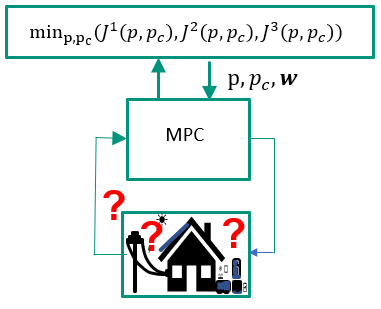}
	\caption{Co-design framework structure}
	\label{fig_codesign}
\end{figure}

The co-design problem, conceptually described in Figure \ref{fig_codesign}, is formulated as the following multi-objective problem
\begin{equation}\label{MOO_co_design}
	\begin{array}{rl}
		&\min_{p,p_c}(J^{(1)}(p,p_c),J^{(2)}(p,p_c), J^{(3)}(p,p_c))\\
		& \textrm{subject to}\\
		& p \in \mathcal{P},  p_c \in \mathcal{P}_c
	\end{array}
\end{equation}
where $\mathcal{P}$ and $\mathcal{P}_c$ describe the feasible spaces for the sizing  and controller parameters, respectively. Note that the costs $J^{(1)}(p,p_c$), $J^{(2)}(p,p_c)$ and $J^{(3)}(p,p_c)$ depend on the-closed loop system performance.
The multi-objective problem (\ref{MOO_co_design}) with conflicting objectives does not have a single solution that simultaneously optimizes each objective, but a set of possible optimal solutions known as the Pareto frontier \cite{Marleretal2004,Ericetal_2019}. A solution is \emph{Pareto optimal} if the improvement in one objective's value degrades some other objective values. 
All Pareto optimal solutions are equally good, and a single choice relies on preference or additional criteria.
In the present contribution, preferences and the proposed approach determine the final optimal solution.

Problem (\ref{MOO_co_design}) is computationally complex since it requires numerous function evaluations consisting of time-consuming simulations to evaluate the closed-loop behaviour under different operating conditions.  
Note that problem (\ref{MOO_co_design})  consists of a bi-level formulation since the design parameters depend on the evaluations of objectives whose explicit expression is unknown. Their values are determined by performing black-box simulations \cite{audetetal2017}.
The derivative information on the objective functions is not available since some of the design parameters included in $p$ and $p_c$ can only assume a finite number of values and the time-varying electricity prices are piece-wise constant.
Moreover, the chosen risk measure can also induce discontinuities in the objective function (e.g. if $\mathcal{R}(\cdot)$ is the $\max$ operator). 

Since problem (\ref{MOO_co_design}) requires costly closed-loop simulations, in the next sections, we propose a simplified formulation capable of giving good choices of the design parameters at a significantly reduced computational burden.

\section{Problem decomposition and importance subsamples}
\label{sec:ProblemDecomposition}
\noindent
The co-design problem requires identifying several significant operating conditions to achieve a reliable system design.
The description of such operating conditions necessitates several years of exogenous data to obtain a set of annual scenarios and a useful uncertainty model of the time-varying exogenous variables.
The multi-objective co-design problem (\ref{MOO_co_design}) with the objective functions defined in (\ref{economic}),(\ref{cl_loop_performance}), and (\ref{MPC_param}) is computationally intractable due to the presence of uncertainty. So a common approach is to consider a limited set of data. 
Unfortunately, a naive choice of subsets of data in most cases gives a solution far from optimal \cite{Hoffmannetal2020,Hilbersetal2020}.

In the present framework, we extend the idea of \emph{importance subsampling}, proposed in \cite{Hilbersetal2020}, to problems presenting dynamics correlating variables at different time instants. The importance subsampling is an approach that extracts fewer observations (subsamples) from long time series through systematic identification of timesteps carrying essential information for the problem under consideration by assigning to each subsample a measure of its importance in realising the problem output.
Since time-varying electricity prices induce a large variability in the costs and the optimal design choices, we determine the importance of a subsample based on the associated optimal cost and the design parameter $p$.
Moreover, since buildings' dynamic behaviour can induce correlations between time instants that are far apart (for example, several days or weeks), the length of the importance subsample might be considered an additional design parameter of the problem.

Even if the multi-objective problem (\ref{MOO_co_design}) operates on reduced data sets, such a problem is still computationally demanding and requires trade-off choices. Therefore, we propose decomposing the problem (\ref{MOO_co_design})  into two problems.
The structure of the problem (\ref{MOO_co_design}) and the physical meaning of the parameters suggest performing the MPC tuning and the sizing separately, but we need to consider if and how the sizing and controller parameters are connected.
In particular, we note that the constraints $\mathcal{P}_c$  and $\mathcal{P}$ on the controller and sizing parameters, respectively, are independent.

The proposed simplified co-design framework aims to mitigate the computational complexity by decoupling multiple objectives and solving multiple optimization problems of reduced size.
The idea is to determine optimal MPC parameters providing good closed-loop performance on a limited significant data set at a moderate computational burden for all the admissible sizing parameters  $\mathcal{P}$.
Consequently, the optimization of the controller considers only objectives directly related to closed-loop performance.
Then, using the determined MPC,  the optimal sizing choice $p^*$ is performed using the economic objective.
The desired optimal $p_c^*$ is computed using the obtained $p^*$.
Validation analysis of the achieved optimal design is possible using data sets not included in the design process.

\subsection{MPC tuning}
The automatic tuning of the MPC controller is given by a trade-off between the objectives $J^{(2)}(p,p_c)$, accounting for the closed loop performance, and $J^{(3)}(p,p_c)$, considering the controller computational complexity, where $p$ is a-priori unknown and consequently considered an uncertain parameter.

The tuning problem $\mathbb{P}^{\mathcal{C}}$ is defined as	
	 
\begin{equation}\label{MOO_auto_tuning}
	\begin{array}{rl}
		&\min_{p_c}\mathcal{R}_p((J^{(2)}(p,p_c), J^{(3)}(p,p_c))\\
		& \textrm{subject to}\\
		& p_c \in \mathcal{P}_c, \;  p \in \mathcal{P},
	\end{array}
\end{equation}
where we have introduced the risk measure $\mathcal{R}_p$ with respect to $p$ since the optimal controller parameters need to be a good choice for all $p\in \mathcal{P}$. In particular, in problem (\ref{MOO_auto_tuning}), we use as a risk measure the $\max$ operator to guarantee the performances on the Pareto frontier for all possible system configurations described by the compact set  $\mathcal{P}$. 
The optimal $p_c$, denoted as $p^*_c$, is a preferred choice determined by a compromise between closed-loop performance and computational complexity.  
The formulation of the problem $\mathbb{P}^{\mathcal{C}}$ can be further simplified since the evaluation of the cost function $J^{(2)}(p,p_c)$ can be reliably performed on persistently exciting \cite{WILLEMS2005325} training data sets $\mathbf{w}^r_{[t_0,t_{N^r+N-1}]}$ of limited length.
The persistently exciting requirement on $\mathbf{w}^r_{[t_0,t_{N^r+N-1}]}$ consists of asking for a sufficiently rich data set that guarantees the system controlled by the MPC visits all the operating conditions of interest. 
The training data sets $\mathbf{w}^r_{[t_0,t_{N^r+N-1}]}$ are created from the full time-series making sure of selecting the relevant features of the signals.
Their length must be at least twice as long as the reference MPC prediction horizon, since the performance are evaluated on the closed-loop system across the length of the MPC prediction horizon.
Persistently exciting conditions can be validated on the computed reference trajectory.
Note that tuning on short datasets might depend on the system's initial state. For this reason, the proposed simplified algorithm considers the effect of the initial state.

Let $\mathcal{X}_{p}\subseteq \mathbb{X}$ be the state constraint set depending on the value of the design parameter $p$.
Algorithm 1 details the procedure for performing the MPC tuning for a training set. If the system behaviour is significantly variable, the procedure can be repeated on additional training sets. The final parameter choice verifies that the desired performances are guaranteed in all the investigated operating conditions. 

\begin{algorithm}[!htbp]
\SetAlgoLined
\KwInput{Dataset $\mathbf{w}^r_{[t_0,t_{N^r+N-1}]}$, controller reference parameter $p_c^r$, $m_p$ samples $p^{(i)}\in \mathcal{P}$, $i=1,\ldots,m_p$, and $m_x$ initial states $x_0^{(j,i)} \in \mathcal{X}_{p^{(i)}}$,  $j=1,\ldots, m_x$, $i=1,\ldots,m_p$, and choice preference criterion.}
\KwOutput{Optimal $p^*_c$}
\For{$i = 1,\ldots, m_p$}{
    	 	\begin{enumerate}
			\item Compute $\mathbf{V_{N^r}}(x_0^{(j,i)},t_0,\mathbf{w}^r_{[t_0,t_{N^r+N-1}]},p^{(i)},p_c^r)$, for  $j=1,\ldots, m_x$ 
	        \item Set $x^r_i=x_0^{(j^*,i)}$  where \\
	        $j^* =\arg \max_{j=1,\ldots, m_x} \mathbf{V_{N^r}}(x_0^{(j,i)},t_0,\mathbf{w}^r_{[t_0,t_{N^r+N-1}]},p^{(i)},p_c^r) $ and\\  $\mathbf{V_{N^r}}(x_0,t_0,\mathbf{w}_{[t_0,t_{N^r+N-1}]},p^{(i)},p_c^r))=\mathbf{V_{N^r}}(x^r_i,t_0,\mathbf{w}^r_{[t_0,t_{N^r+N-1}]},p^{(i)},p_c^r)$ in (\ref{cl_loop_performance}).
	        \item  Solve problem $\mathbb{P}^{\mathcal{C}}$ from \eqref{MOO_auto_tuning} imposing $p=p^{(i)}$ and compute  the Pareto frontier $P^f_i$.
      \end{enumerate}
    }
Select $p^*_c$ on  $P^f_i$, $\forall \; i=1,\ldots,m_p$ accordingly to another quantitative or qualitative preference criterion \cite{zhu2021cglisp}.
\caption{MPC tuning \label{alg1}}
\end{algorithm}

Note that when the states are related to the stored energy in the system, the number $m_x$  of needed initial conditions is often small.
Indeed the most demanding operating conditions for energy storage often sit on the extremes given by minimal or maximal low energy levels, which require higher control effort to balance internal energy requirements.
In addition, the optimal $p^*_c$ also gives information on the length of the importance subsample, which cannot be shorter than twice the length of the MPC prediction horizon.

\subsection{Importance subsample}\label{sub:Importance}

\noindent 
Classical data reduction approaches use individual years or cluster data into representative days and lead to significant errors in estimates of optimal system design due to data omissions affecting the output of the problem since they neglect how the problem depends on the data \cite{HILBERS2019113114}.
Conversely, the importance subsampling approach selects and groups subsamples according to their effect on the problem output, as discussed in \cite{HILBERS2019113114} for models without interdependence between the sampled data.

\subsubsection{Subsample definition }
In the present contribution, the importance of a subsample is evaluated by optimizing the system investment and operation cost computed, considering the closed-loop operation across a short subsample of weather and electricity prices. Note that the subsample length, as pointed out in the previous subsection, needs to be longer than twice the MPC prediction horizon to consider the correlation between time instants through the MPC prediction horizon on the closed-loop performance. The operation cost of the closed-loop system across the short subsample is weighted by $R_h$ according to the sub-sample length to estimate the annual operational cost, assuming that the considered operating conditions repeat across the year.

Let ${\mathcal{S}}:=\{\mathbf{S^{(1)}}, \mathbf{S^{(2)}}, \ldots \mathbf{S^{(m)}}\} $ be a collection of $m$ subsamples of $\mathbf{w}_{[t_0,T]}$ where $T$ accounts for the full length of the time-series. Each $\mathbf{S^{(h)}}$ has $N_h+N-1$ samples where $N_h$ is the simulation length.
The problem, denoted as $\mathbb{P}_{h}^{\mathcal{S}}$, evaluating the importance of the subsample $\mathbf{S^{(h)}}$ is defined as
\begin{equation}\label{importance_pb}
	\begin{array}{rl}
	V^*_h(\hat{x}_h,p_c^*)= &\min_{p} 
	R_h \mathbf{1'_{N_h}}\mathbf{V_{N_h}}(\hat{x}_h,t^{(h)}_0,\mathbf{S^{(h)}},p,p_c^*)+V_I(p)\\
		& \textrm{subject to  }  p \in \mathcal{P}, 
	\end{array}
\end{equation}	
where $\mathbf{1_{N_h}}$ is a vector of ones of length $N_h$, $t^{(h)}_0$ is the initial time of $\mathbf{S^{(h)}}$
and $\hat{x}_h$ is the assigned initial state.
The choice of $\hat{x}_h$ should represent a condition providing a degree of robustness according to a risk measure. A natural simplifying choice for energy-related problems is considering the initial state returning the $\mathbb{P}_{h}^{\mathcal{S}}$ with a higher cost.

If the specific problem instance is solvable for a prediction horizon covering the whole subsample, a meaningful choice is to consider the initial state a decision variable and impose an equality constraint with the state at the end of the subsample.   
The selection of subsamples requires the optimal solution of $\mathbb{P}_{h}^{\mathcal{S}}$ for $h=1, \ldots, m $ and the definition of criteria to determine their importance. The criterion is a problem-dependent choice, and the most common choices consider the optimal cost 	$V^*_h(\hat{x}_h,p_c^*)$ as discussed in \cite{HILBERS2019113114,Hilbersetal2020}.
However, volatile electricity time-varying prices make the optimal cost as a single criterion not always adequate. Therefore, we propose choosing the importance subsample using the optimal cost and the design parameters $p$. This because, independently of the achievable prices, they represent the design choice of interest for the problem under consideration.
Algorithm \ref{alg:ImpSub} defines the procedure to select the representative subsamples. 

\begin{algorithm}[!htbp]
\SetAlgoLined
\KwInput{Get extensive time-series of exogenous signals, 
	break the dataset into $m$ subsamples $\mathbf{S^{(h)}}$  for $h \in H=\{1,\ldots,m \}$, set initial state conditions $\hat{x}_h$, controller parameters $p_c^*$, maximum number of clusters $k_{\max}$,  choose clustering technique, distance measure between points in the cluster and their representative and the desired maximum distance $d_{\max}$}
\KwOutput{Number of clusters $n_c$, clusters' representatives and number of data-points $\nu_i$ in the cluster $i$ for $i=1, \ldots,n_c$ }
 Determine data-points $y_h:=(p^*_h,V^*_h(\hat{x}_h,p_c^*))$  solving problems $\mathbb{P}_{h}^{\mathcal{S}}$ for $h=1,\ldots m$;\\
 $ k \gets 2 $ \\
  \While{$k \leq  k_{\max}$}{
  Partition the $m$ points into $k$ groups $\mathcal{G}_i $ for $i=1,\ldots,k$, accordingly to the chosen criterion,  such that $\mathcal{G}_i\bigcap \mathcal{G}_j = \emptyset $ for all $i \neq j$;  determine representatives $C_i$ for each cluster $\mathcal{G}_i$ for $i=1, \ldots,k$;\\
  Compute $D_{C_i}= \max_{y_h \in \mathcal{G}_i}d(y_h,C_i) $  for $i \in \{1, \ldots, k\}$ \\
  \eIf{$ \max_i D_{C_i} \geq d_{\max}$}
{
    $ k\gets k+1$\;
}{
  Exit
}
}
Set $n_c:=k$, 
    $\mathcal{H}_c=\{h \; | \; \exists i \; : \; y_h= C_i \} $ and $\nu_i:=|\mathcal{G}_i|$  for $i=1,\ldots,k$ 
\caption{Importance Subsamples \label{alg:ImpSub}}
\end{algorithm}
\noindent
In Algorithm \ref{alg:ImpSub}, $|\mathcal{G}_i|$ denotes the cardinality of the set $\mathcal{G}_i$ and  the function $d(\cdot)$ computes the distance between points.Various distance measures and methods appear in the literature to evaluate the quality of a cluster \cite{kaufmanetal2009,Madhulatha2012}.
A commonly used distance is the within-cluster sum of squares, which estimates cluster tightness and accounts for the variance of the data, but different distance measure choices are possible.

\subsubsection{Clustering algorithms}

The most popular partitioning algorithms are known as k-means and k-medoids, and they minimize the sum over each cluster of the squared distance between the selected cluster centre and candidate points of the cluster. The k-medoid algorithm has the advantage that the centre is an element of the cluster and it is more robust to outliers and noise, as discussed in \cite{Jin2010}. 

The choice of the number of clusters is not, in general, only determined by the desired clustering resolution, but is a separate task from the clustering. 

The shape of the clusters and the scale of the distribution of data points are important factors, as the case study shows, and the correct choice of $n_c$ is often unclear.

We finally detail the framework to perform the system co-design, integrating the previously introduced algorithms. 
In particular, the design parameter $p^*$ is a solution to the  following problem $\mathbb{P}^{\mathcal{CD}}$, which uses the information obtained by running Algorithms \ref{alg1} and \ref{alg:ImpSub}. The objective now is:
\begin{equation}\label{size_design} 
	\min_{p\in \mathcal{P}}
	 \mathcal{R}\left(\sum_{i \in \mathcal{H}_c}  \nu_i \mathbf{1'_{N_i}}\mathbf{V_{N_i}}(\hat{x}_i,t^{(i)},\mathbf{S^{(i)}},p,p_c^*)\right) +V_I(p)
\end{equation}
The problem $\mathbb{P}^{\mathcal{CD}}$ provides a design parameter $p^*$ that has a degree of robustness that is the result of a compromise between computational complexity and modelling accuracy.
The full co-design procedure is synthesized in Algorithm \ref{alg:co_design}. \\

\begin{algorithm}[!htbp]
\SetAlgoLined
\KwInput{System data, input parameters for Algorithms \ref{alg1} and \ref{alg:ImpSub}.}
\KwOutput{Optimal $(p^*,p_c^*)$}
\textbf{MPC tuning }: Execute Algorithm 1 and determine controller parameters $p_c^*$   \\
\textbf{Importance subsamples selection}: Execute Algorithm 2 using $p_c^*$ and determine number of clusters  $n_c$, their representatives identified by  $\mathcal{H}_c$ and the corresponding weight $\nu_i$ for $i=1,\ldots,n_c$. \\
Compute the optimal $p^*$ solving problem  $\mathbb{P}^{\mathcal{CD}}$.\\
Validate the performance of $p^*$ on different datasets. If the achieved performance is not satisfactory, add the subsample presenting unsatisfactory performance, update the number of clusters and the cluster parameters for $\mathbb{P}^{\mathcal{CD}}$  and go to 3.\\
Update $p_c^*$ running algorithm 1 with $p=p^*$ and set as reference $p_c^r$ the optimal parameter $p_c^*$ used for the co-design.\\
Return $p_c^*$ and $p^*$.
 \caption{System co-design \label{alg:co_design}}
\end{algorithm}
\noindent

\section{Case study: Residential building co-design}
\label{sec:CaseStudy}
The importance and effectiveness of the proposed co-design framework are illustrated on 
a residential building contributing to a grid with time-varying energy prices. 

\subsection{Dynamic optimization under uncertainty}
The thermal dynamics of a three-bedroom dwelling with a high insulation level is modelled by adopting a single-zone lumped-capacitance method \cite{Hazyuk2012}. The building is equipped with electrically driven heat pumps (HP), providing temperature regulation. We consider the option of installing photovoltaic panels (PV) and rechargeable lithium batteries. The surface area $S^{PV}$ covered by the PV panels and the battery capacity $S^B$ are the designing parameters defining $p:=[S^{PV}, \;  S^B ]$. The system state $x(t):= [T(t), \; SoC(t) ]'$  includes the building internal temperature $T(t)$  and the battery state of charge  $SoC(t)$.

The input  $u(t):=[u^{eH}(t), \; u^{CeH}(t), \; u^{dch}(t), \; u^{ch}(t), \; u^{b}(t), \; u^{s}(t) ]'$  consists of the electricity power $u^{eH}(t)$ and $u^{CeH}(t)$  consumed by the heating and cooling pumps, respectively, the battery charging $u^{ch}(t)$  and discharging  $u^{dch}(t)$  rates and the bought $u^{b}(t)$ and sold $u^{s}(t)$ power. The uncertain exogenous vector  $w(t):=[T^e(t), \; I(t), \; c^{el}(t), \;  c^{em}(t)]$  considers the external temperature $T^e(t)$, solar irradiance $I(t)$, electricity prices  $c^{el}(t)$ and carbon emissions $c^{em}(t)$.
The dynamics of the building are as follows   
 \begin{equation}\label{building_dynamic}
	\begin{array}{l}
		\left[\begin{array}{c}
			\dot{T}(t) \\
	        \dot{SoC}(t)\\
		\end{array} \right]=
		\left[\begin{array}{cc}
			-(U A+\rho_{air} V C_{air}^p n_{ac})/C_{build} &  0 \\
	         0 &  0 \\
		\end{array} \right]\left[\begin{array}{c}
			T(t)\\
			SoC(t)\\
		\end{array} \right] +\\
		\left[\begin{array}{cccc}
			COP(T^e(t))/C_{build} & -COP_{cool}/C_{build} & 0 & 0 \\
		   0  &  0 &  -1/\eta^{ds} & \eta^{ch} \\
		\end{array} \right]\left[\begin{array}{c}
			u^{eH}(t)  \\
		    u^{CeH}(t) \\
			u^{dch}(t) \\
			u^{ch}(t)
		\end{array} \right]+\xi(t),
	\end{array}
\end{equation}
where 
\begin{equation}
	\xi(t)=
	\left[ \begin{array}{l}
		T^e(t)(U A+\rho_{air} V C_{air}^p n_{ac})/C_{build} \\
		 0
	\end{array} \right]
\end{equation}
where $\text{COP}(T^e(t))=m_{\text{COP}} (T^e(t)-7) +3$. The values of all parameters are given in Table \ref{Build_param}.             
\begin{table}[tb]
	\caption{Dwelling parameters}
	\label{Build_param}
	\centering
	\begin{tabular}{l c c c}
		\hline     
		$\mathbf{Description}$ &  $\mathbf{Parameter}$ & $\mathbf{Value}$ & $\mathbf{Unit}$ \\ 
		\hline
		Average U-value  & $U$ & $0.93195$ &  W/(m$^2$ K) \\
		Wall surface area & $A$ &	$82.06959707$ &	m$^2$ \\		
		Air density	 & $\rho_{air}$ & $1.225$ & kg/m$^3$    \\
		Building volume & $V$ & $224.05$  & m$^3$ \\
		Air heat capacity & $C_{air}^p$ & $1.005$ & kJ/(kg K) \\
		Air changes per hour & $n_{ac}$ &  $1$ & h$^{-1}$ \\	
		Building thermal mass & $C_{build}$ & $15286.6114$ &  kJ/K \\
		Floor surface area	&  $S_F$ & $89.62$ & m$^2$ \\
		HP electricity bound	  & $\overline{u}^{eH}$   &  $4$  &	kW  \\
		CP electricity bound &  $\overline{u}^{ceH}$   &  $6$  &	kW  \\
		HP capacity	        & $\overline{Q}^{HP}$   &  $6$  &	kW  \\
		\hline
	\end{tabular}
\end{table}
The imposed input and state constraints are
\begin{align}
	& \underline{T}(t) \leq T(t) \leq \overline{T}(t) \label{comfort}\\
	& 0 \leq SoC(t) \leq   S^B \leq \overline{SoC} \label{bound_SB}  \\    
	& 0 \leq u^{dch}(t), u^{ch}(t) \leq  S^B /T_{ds}  \label{battery_B}\\
	& P^{PV}(t)=\theta_1(1+\theta_2 I(t) + \theta_2 T^e(t))I(t) S^{PV}  \label{pv_power}\\ 
	&	u^b(t)-u^s(t)+u^{dch}(t)-u^{ch}(t)+P^{PV}(t)=u^{eH}(t)+u^{CeH}(t)\\
	&	0 \leq  u^b(t) \leq \overline{u}^b, \;  0 \leq  u^s(t) \leq \overline{u}^s \\
	& 0 \leq  u^{eH}(t) \leq \overline{u}^{eH}, \; 0 \leq  u^{CeH}(t) \leq \overline{u}^{CeH} \\
	&\text{COP}(T_t^e) u_t^{eH} \leq  \overline{Q}^{HP}\\
	&   0 \leq  S^{PV} \leq  S_F,    \label{QoST}
\end{align}
where $\underline{T}(t)$ and $\overline{T}(t)$ define thermal comfort limits according to standards \cite{CIBSE}, $T_{ds}$ is the number of hours required to fully discharge the battery at the maximum rate and $P^{PV}(t)$ is the power produced by the PV panels. 
All the parameters used in the study are reported in Table \ref{asset_param}.

The nonlinear function (\ref{pv_power}) is a good model of the maximum power generated by PV panels \cite{PVUSA,PV_model}, while the 
operating limits refer to the design specs for the multi-crystalline JAP6 4BB module range produced by JA \cite{solar_2020}. 
\begin{table}[tb]
	\caption{Problem Parameters}
	\label{asset_param}
	\centering
	\begin{tabular}{c c c c}
		\hline      
		$\mathbf{Description}$ &  $\mathbf{Parameter}$ & $\mathbf{Value}$ & $\mathbf{Unit}$ \\ 
		\hline
		HP COP slope & $m_{\text{COP}}$ & $0.067$ &  ${\;}^{\circ} $C\\
		CP  COP  & $\text{COP}_{cool}$ & $0.7$ &  -  \\
		Battery charging & $\eta^{ch}$ & $0.88$ & - \\
		Battery discharging & $\eta^{ds}$ & $0.88$ & - \\
		Discharging hours   & $T_{ds}$   & $2$ & h \\
		Bought power bound & $\overline{u}^b$ &  $30$ & kW \\
		Sold power bound & $\overline{u}^s$ &  $30$ & kW \\
		Max battery size & $\overline{SoC}$ &  $60$ & kWh \\
		Power/($I_r$)  gain &  $\theta_1$ & $0.12$ & kW/m${}^2$\\
		Power/($I_r$)  correction &  $\theta_2$ &   $-1.345e^{-4}$ & - \\
		Power/($T$ $I_r$)  correction  &  $\theta_2$ &   $-3.25e^{-3}$ & - \\
		Carbon price     &  $c_{CO_2} $ &  $100$    &  £/(ton CO2e)\\
	    Battery lifespan	&    $y$  &  $15$    &  years\\
	    PV lifespan        &    $y$  &  $30$    &  years\\
	    Battery CAPEX    & $C_B$    &  $460$   &  £/kWh\\
	    PV CAPEX        &   $C_{PV}$     & $325$      &  £/$\textrm{m}^2$\\
	     Interest rate   &   $i_r$        & 2\%      &   -\\
		\hline
	\end{tabular}
\end{table}

The stage cost used by the EMPC depends on the electricity prices  $$\ell(x(t),u(t),w(t),t,p) :=  c^{el}(t)u^{b}(t)- 0.9c^{el}(t) u^s(t)+c_{CO_2} c^{em}(t)u^{b}(t).$$
The time-varying electricity prices $c^{el}(t)$ are piece-wise constant with $15$ minutes resolution. The price data assumes the Octopus Agile tariff pricing mechanism \cite{Octopus} using the Market Index Price and data from \cite{elexon}. Weather data have been obtained from the Centre for Environmental Analysis (CEDA) archive \cite{ceda}. The grid CO2 intensity is based on data from the Carbon Intensity API developed by the ESO National Grid \cite{carbon}.  

The expenditures  $V_I(p)=c_B S^B+c_{PV} S^{PV}$ faced to buy the technologies use  annualised capital costs $c_B$ and $c_{PV}$ 
computed by dividing the capital cost (CAPEX) by the ``present value of annuity factor"
\[
a_{y,r}=\frac{1-\frac{1}{(1+r)^y}}{r}
\]
considering the technology lifespan and the interest rate reported in Table \ref{asset_param}. 
The technologies come in units of $1 \,\textrm{kWh}$ for the battery capacity and $1.68 \,\textrm{m}^2$ for the PV panel dimension.

\subsection{Numerical results}

\noindent The presented studies highlight the importance of performing a co-design in an integrated fashion, including uncertainty due to different possible operating conditions and the importance of its various components. 

\subsubsection{Numerical setup}
The adopted risk measure in the cost (\ref{economic}) is the expectation over $11$ scenarios corresponding to the available data. The framework has been implemented in MATLAB. In particular, the EMPC formulation realising the closed-loop simulation uses ICLOCS2.5  \cite{iclocs2} with the solver IPOPT \cite{ipopt} while NOMAD \cite{Le2011a} has been used to solve the black-box optimization problems.
The adopted transcription method is explicit Euler, since it allows input discontinuities. The clusters were computed using the function k-medoid implemented in MATLAB. The studies were performed on a server with an AMD EPYC 7443 24-Core processor and running Windows Server  2019. 

\subsubsection{Deterministic and robust solution}
In the first study, we compare the technology sizing using multiple scenarios against the deterministic case to demonstrate the importance of considering different operating conditions. For this study we have set $T^{(d)}_k=T^{(s)}=15\,$min  for all $k=0,1,\ldots, N$ and $t_f=24 \,$h.

Table \ref{tab:deterministic} reports $11$  optimal technology designs according to a deterministic formulation using the information of a single year.
For each year, the mean cost in Table \ref{tab:deterministic}  corresponds to the annual mean cost achieved by applying the obtained optimal deterministic design across the $11$ years with the different realizations of the exogenous signals. 
The results show a large variability in the size choice depending on the year under consideration. The mean cost over the considered $11$ years is much higher than the optimal cost of the planning phase. \\
\begin{table}[!tb]
	\centering
	\caption{Deterministic sizing for $T^{(d)}_k=T^{(s)}=15$min  for all $k=0,1,\ldots, N$ and $t_f=24$h.}
	\label{tab:deterministic}%
	\begin{tabular}{ccccc}
		\toprule
		\multicolumn{1}{p{4.055em}}{Data Year} & \multicolumn{1}{p{4.055em}}{Battery (kWh)} & \multicolumn{1}{p{4.055em}}{PV area (m${}^2$)} & \multicolumn{1}{p{5.665em}}{Optimal Cost  (£/year)} & \multicolumn{1}{p{6.11em}}{Mean Cost (£/year)} \\
		\midrule
		2008  & 60    & 89.0  & -563.6 & 415.3 \\
		2009  & 60    & 0.0   & 248.1 & 491.0 \\
		2010  & 0     & 3.4   & 451.1 & 386.1 \\
		2011  & 0     & 89.0  & 261.3 & 379.5 \\
		2012  & 0     & 8.4   & 411.8 & 381.6 \\
		2013  & 51    & 89.0  & 350.4 & 393.9 \\
		2014  & 0     & 1.7   & 316.4 & 388.1 \\
		2015  & 0     & 1.7   & 309.6 & 388.1 \\
		2016  & 60    & 5.6   & 237.6 & 491.0 \\
		2017  & 0     & 0.0   & 382.4 & 390.5 \\
		2018  & 9     & 89.0  & 208.8 & 361.2 \\
		\bottomrule
	\end{tabular}%
\end{table}%
\indent
Conversely, the solution returned by the robust formulation, reported in Table \ref{tab:robustcodesign},  is substantially different from the solutions to the deterministic problem.
In Table \ref{tab:robustcodesign}, the cost denoted as ``Effective mean cost" is the optimal cost achieved by solving the co-design problem (\ref{MOO_co_design}) using the mean as the risk measure and optimizing the parameters $p$ or fixing the value of $p$ to the solution returned by the optimization problem  (\ref{importance_pb}).
The optimal cost obtained by solving (\ref{importance_pb}) 
for different cluster choices appears in Table \ref{tab:robustcodesign} as ``Estimated mean cost". 
Table \ref{tab:robustcodesign} also compares the optimal sizing and the computational times obtained by solving the co-design problem (\ref{MOO_co_design})  warm started with the value $p=[9, ~44]$ against the robust co-design problem (\ref{importance_pb}) by exploiting parallel computation.
Remarkably robust formulations achieve a better cost in all cases, up to a reduction of about $30\%$ in some cases. 
Moreover, the decomposed robust formulation (\ref{importance_pb}) returns optimal solutions with performance comparable to the optimal solution to (\ref{MOO_co_design}) with substantially reduced computational time up to about $90\%$.
\begin{table}[tb]
	\centering
	\caption{Technology sizing using robust formulations with pre-defined EMPC parameters $T^{(d)}_k=T^{(s)}=15$min  for all $k=0,1,\ldots, N$ and $t_f=24$h.}
	\label{tab:robustcodesign}%
	\begin{tabular}{ccccccc}
		\toprule
		\multicolumn{1}{p{4.355em}}{Co-design Problem (Pb)}   & \multicolumn{1}{p{3.555em}}{Scaling in
		Sub-sampling}   &  \multicolumn{1}{p{3.055em}}{Battery (kWh)} & \multicolumn{1}{p{3.055em}}{PV area (m${}^2$)} & \multicolumn{1}{p{3.665em}}{Effective mean cost (£/year)} &
		\multicolumn{1}{p{3.655em}}{Estimated mean cost (£/year)} 
		& \multicolumn{1}{p{4.11em}}{Comp. time (days)} \\
		\midrule
		Pb (\ref{MOO_co_design}) & - & 16   & 89.0 & 347.1 & - & 45 \\
		Pb (\ref{size_design}) $n_c=574$ & no & 13  &  89.0 & 347.6 &  356.7 & 30 \\
		Pb (\ref{size_design}) $n_c=50$ & no  & 23  &  89.0 & 349.3 &  378.2 & 4 \\
		Pb (\ref{size_design}) $n_c=50$ & yes & 14  &  89.0 & 347.3 &  303.4 & 2.8 \\
	\bottomrule
	\end{tabular}

\end{table}%

The problem considering a parallel implementation over $574$ sub-samples returns an optimal solution close to the original problem in terms of technology sizes and optimal cost. In particular, the solution to problem (\ref{size_design}), considering $574$ week-long sub-samples, returns a mean cost of $356.7\pounds/$year which is a close estimate of the effective cost of $347.6\pounds/$year  achieved by the optimization problem (\ref{MOO_co_design}) using a battery of $13 \,$kWh and $89.0 \,$m${}^2$ of PV panels. The difference in the optimal solution is due to the building's initial state condition.

The optimal solution to the problem that simulates the closed-loop for short periods is very sensitive to the choice of the initial condition.
The initial condition describes the energy available in the building at the beginning of the period, and its value affects the overall control outcome. The performed studies consider an empty battery and the temperature at its minimum value at the beginning of each sub-sample. The used initial state describes the worst condition in the winter season and adds a degree of robustness to the design process. The computation time achieved by the parallel implementation, including all the available data, is lower but still substantial.

Table \ref{tab:robustcodesign} also reports the optimal design applying the importance sub-sample technique described in Subsection \ref{sub:Importance}. 
The computational time is about $10$ times lower compared to problem (\ref{MOO_co_design}), with a deterioration in the performance of only $0.6\%$.

\begin{figure}[!tb]
    	\centering
    	\includegraphics[width=0.8\textwidth]{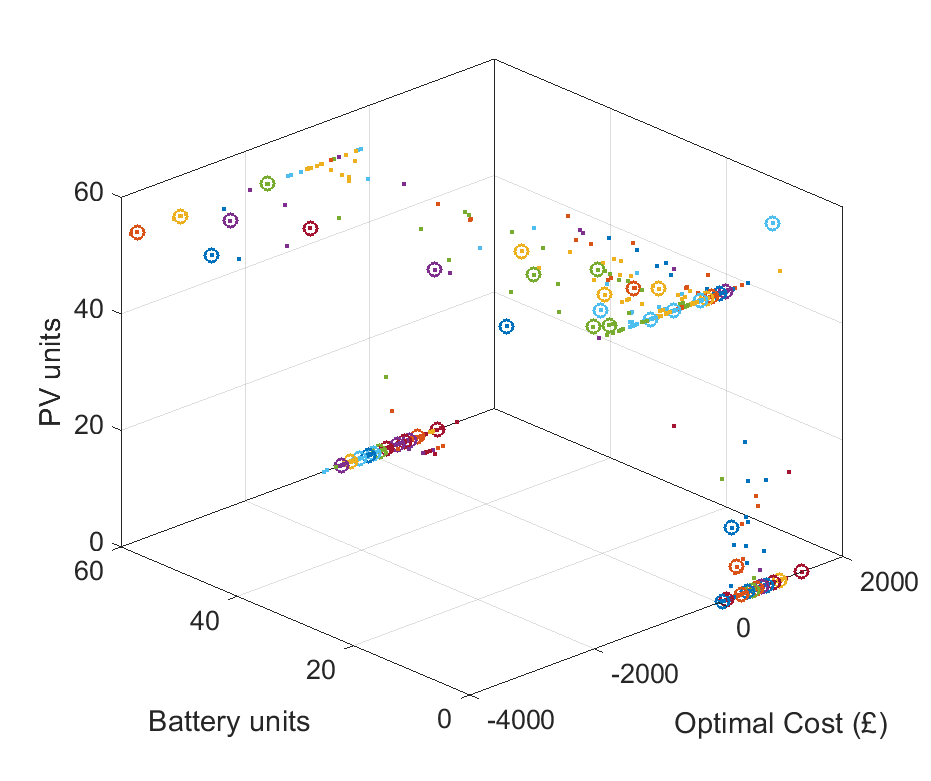}
    	\caption{Data points (coloured dots) consisting in the optimal cost and technologies' size solution to (\ref{importance_pb}) clustered in $50$ groups differentiated by colours. Circles indicates the clusters' centroids.}
    	\label{Fig_cluster_noscal}	
 \end{figure}

\subsubsection{Clustering performance}
The clusters are determined on data points considering the optimal costs and sizes computed by solving problem (\ref{importance_pb}). The output of Algorithm \ref{alg:ImpSub}  considering $50$ groups is reported in Figure \ref{Fig_cluster_noscal}. 

\begin{figure}[!tb]
    \centering
    	\includegraphics[width=0.5\textwidth]{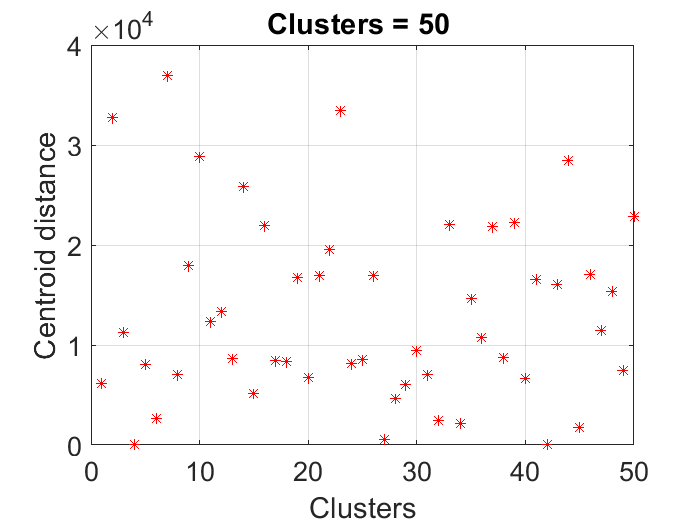}
    	\caption{Within-cluster sums of point-to-medoid distances of the clusters in Figure \ref{Fig_cluster_noscal}}
    	\label{Fig_centroid_dis_k50}
\end{figure}

\begin{figure}[!tb]
    \begin{subfigure}{0.5\textwidth}
    	\centering
    	\includegraphics[width=\textwidth]{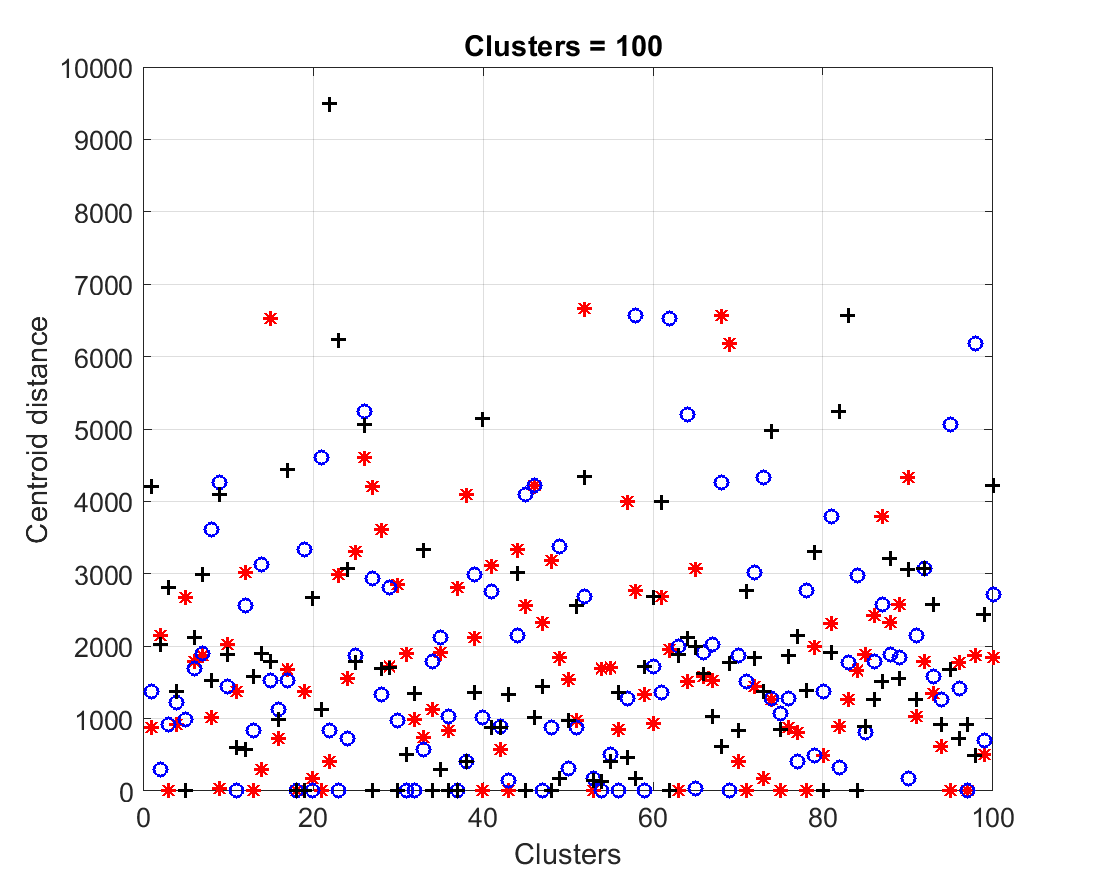}
    	\caption{Distances for $100$ clusters. Different markers indicate the outcome of different runs of the clustering algorithm.}
    	\label{Fig_Cluster_100}	
    \end{subfigure}
    \begin{subfigure}{0.45\textwidth}
    	\centering
    	\includegraphics[width=\textwidth]{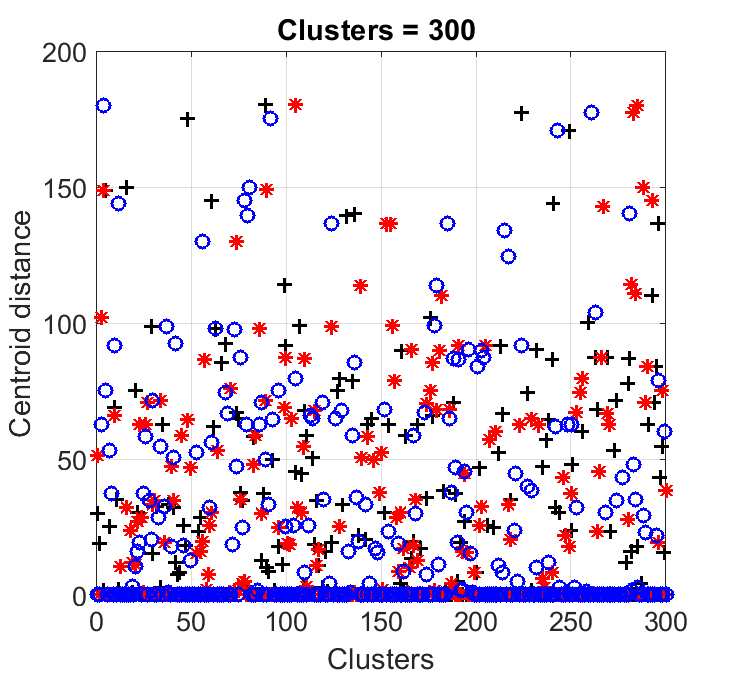}
    	\caption{Distances for $300$ clusters. Different markers indicate the outcome of different runs of the clustering algorithm.}
    	\label{Fig_Cluster300}	
    \end{subfigure}
    	\caption{Within-cluster sums of point-to-medoid distances for different numbers of clusters' choices}\label{fig:cluster2}
\end{figure}

Note the large variability in the values of the optimal costs, which gives a large sum of distances  between a data point and the center of its cluster, as shown in Figure  \ref{Fig_centroid_dis_k50}.
In particular, Figure \ref{Fig_centroid_dis_k50} reports the sum of the Euclidean distances from the centers, called centroids, considering $50$ clusters. 
The cluster spread can be only reduced by substantially increasing the number of clusters, as shown in Figure \ref{fig:cluster2}.

Moreover, the clustering algorithm returns different outputs every time the algorithm is executed, as shown in Figure \ref{fig:cluster2},  depending on the choice of the first centroid.   
Even if the clustering routine uses the  K-means++ algorithm \cite{Arthuretal2007} to avoid the problem of sensitivity to the initialization, different runs of the clustering algorithm return  different centroids with substantially different sizing solutions. Also, note that in Figure \ref{Fig_cluster_noscal} the selected centroids do not adequately represent all data points due to the dominance of the high values of the cost in the clustering procedure.
Consequently, the clustering has been performed on data points with a scaled cost assuming values in  $[-60, ~ 60]$ to assign the same importance to all the quantities and reduce sensitivity to the initialisation.
The centroids, obtained by re-scaling the data points, cover more uniformly the space reducing the sensitivity of the sizing to the specific clustering output.   
The scaling essentially defines the importance of a quantity in the clustering process, and regularises data considering their semantic meaning.

\begin{figure}[!tb]
    \begin{subfigure}{0.45\textwidth}
    	\centering
       \includegraphics[width=\textwidth]{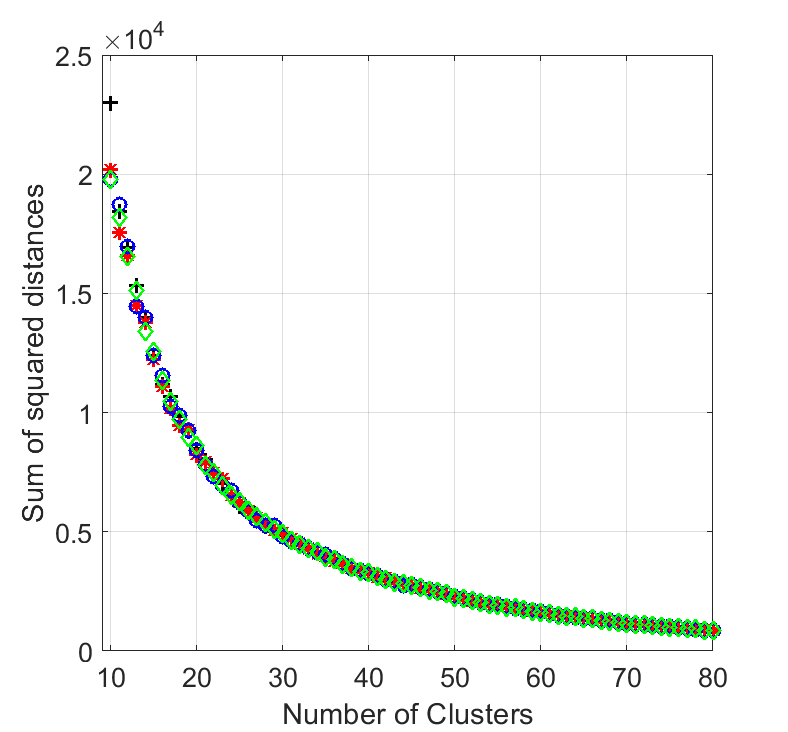}
    	\caption{Sum of errors as functions of the number of clusters with scaling for multiple experiments }
    	\label{Fig_elbow_scal}	
    	
    \end{subfigure}
    \begin{subfigure}{0.45\textwidth}
    	\centering
    	\includegraphics[width=\textwidth]{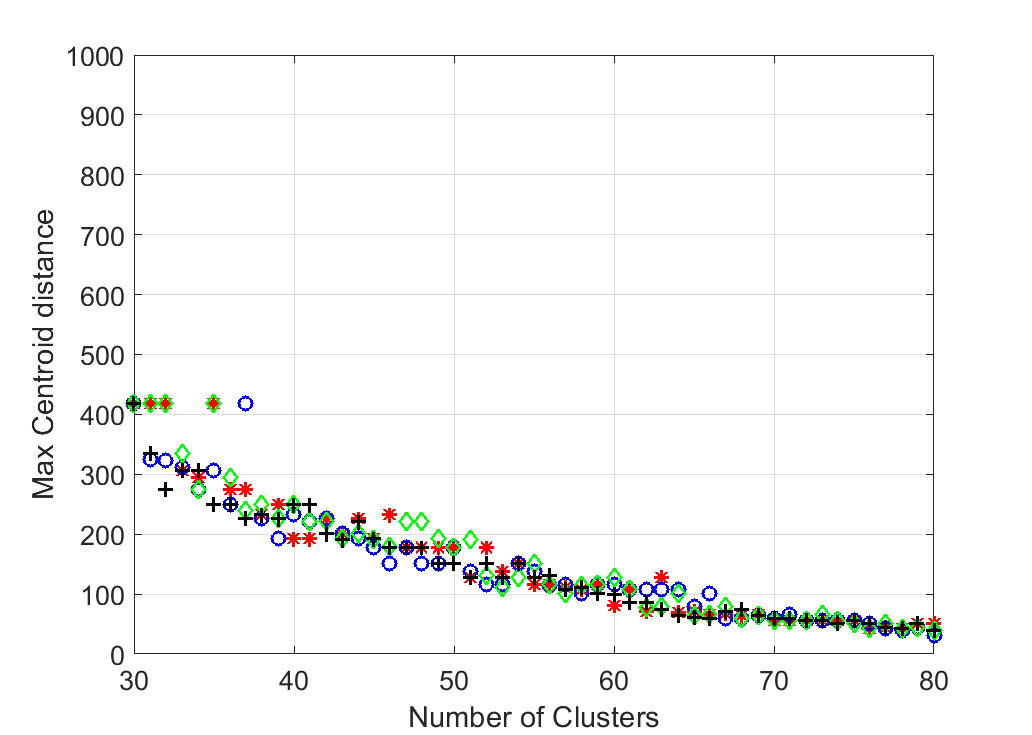}
    	\caption{Maximum of errors as functions of the number of clusters with scaling for multiple experiments}
    	\label{Fig_numcluster_scal}	
    \end{subfigure}
\end{figure}

The choice of the number of clusters is not straightforward. The application of classical empirical techniques such as the Elbow and the Silhouette methods \cite{cluster_over} fails to give a clear indication of an optimal choice of the number of clusters. Figure 
\ref{Fig_elbow_scal} show the sum of the
squared distances.  
Commonly, the optimal number of clusters is identified as the value associated with a slope change (Elbow) in the sum of squared distance as a function of the number of clusters.
However, Figure \ref{Fig_elbow_scal} does not show any slope change and it is inconclusive. 
Instead, the maximum of the summed point-to-centroid distances in Figure \ref{Fig_numcluster_scal} highlights the variability of the clustering outcome.  
Figure \ref{Fig_numcluster_scal} suggests that a clusters' number larger than $50$ reduces the uncertainty related to the specific cluster realisation. 

\subsubsection{Technology-specific MPC tuning}

An interesting outcome of the case studies is the observation that the choice of MPC parameters depends on the technology size. The MPC tuning of the sampling time, discretization step and prediction horizon, considering the constraints induced by the discontinuities at the changes in the electricity prices, uses the discrete variables $n_s$, $n_x$ and $n_f$ characterising the variables of interest as follows
\begin{align}
& T^{(s)}:=n_s T^{(d)} \\
& \delta_T:=n_x T^{(s)} \\
& t_f= n_f  
\end{align}
where $\delta_T$ denotes the time interval at which the change in the price occurs. A resolution of an hour in the prediction horizon is reasonable for the co-design problem as a first approximation to limit the computational burden.
The discretization step is implicitly defined as 
$$
 T^{(d)}=\frac{\delta_T}{n_x n_s}.
$$
The physical limitations on the tuning parameters induce the following bounds on the decision variables \begin{align}
&  1\leq n_s \leq  \frac{\delta_T}{ \underline{T}^{(d)}} \\
&  1\leq n_x \leq  \frac{\delta_T}{ \underline{T}^{(d)}}\\
&  n_s n_x \leq  \frac{\delta_T}{ \underline{T}^{(d)}}
\end{align}
where $\underline{T}^{(d)}$ denotes the lower bound of the discretization step.
The performed studies consider as objective ($\ref{MPC_param}$)  $Q(p_c):=n_f-1/(4n_x)-1./(4 n_x n_s)$ with $p_c=[n_s, n_x, n_f]$, corresponding to minimising the prediction horizon and maximising the sampling time and the discretization step.  
The value of $\underline{T}^{(d)}$ is $5$ minutes, and the reference trajectory used in the objective $J^{(2)}(p,p_c)$ considers $T^{(d)}=T^{(s)} =5$ min and $t_f=3$ days.
The tuning algorithm runs the closed-loop MPC considering a whole week. 

\begin{figure}[!tb]
   \centering
    	\includegraphics[width=0.6\textwidth]{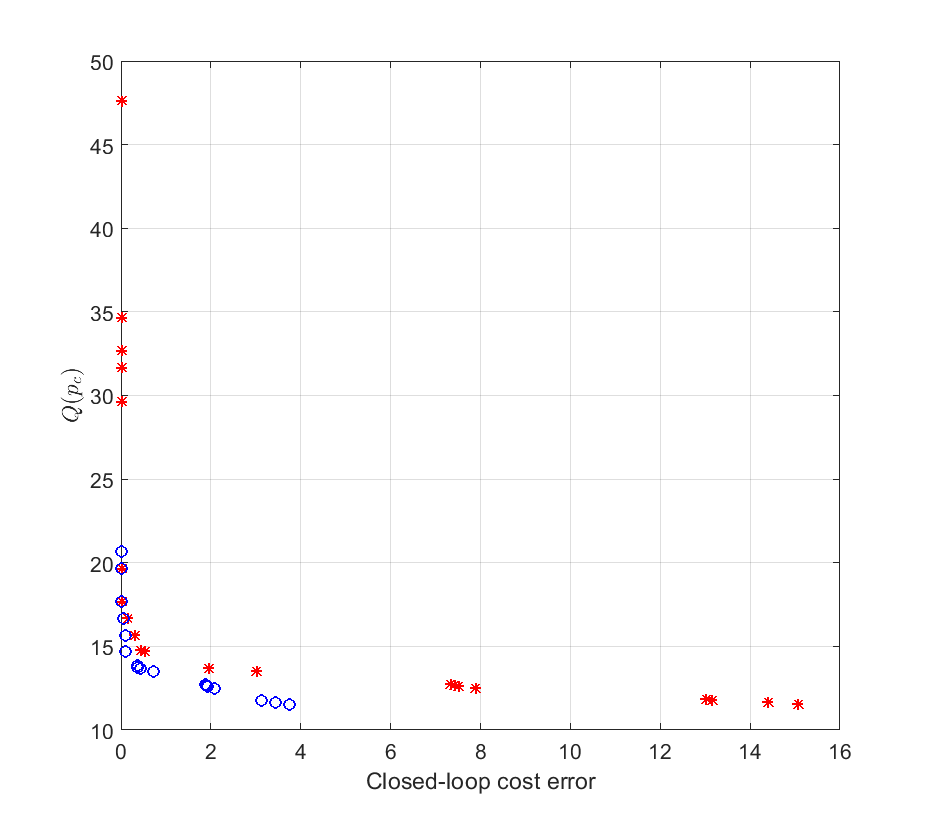}
    \caption{MPC tuning - Pareto front for the maximum technology size (red *) and for half technology size  (blue o)}
    	\label{Fig_paretopc}	
\end{figure}

\begin{figure}[!tb]
   \centering
    	\includegraphics[width=0.6\textwidth]{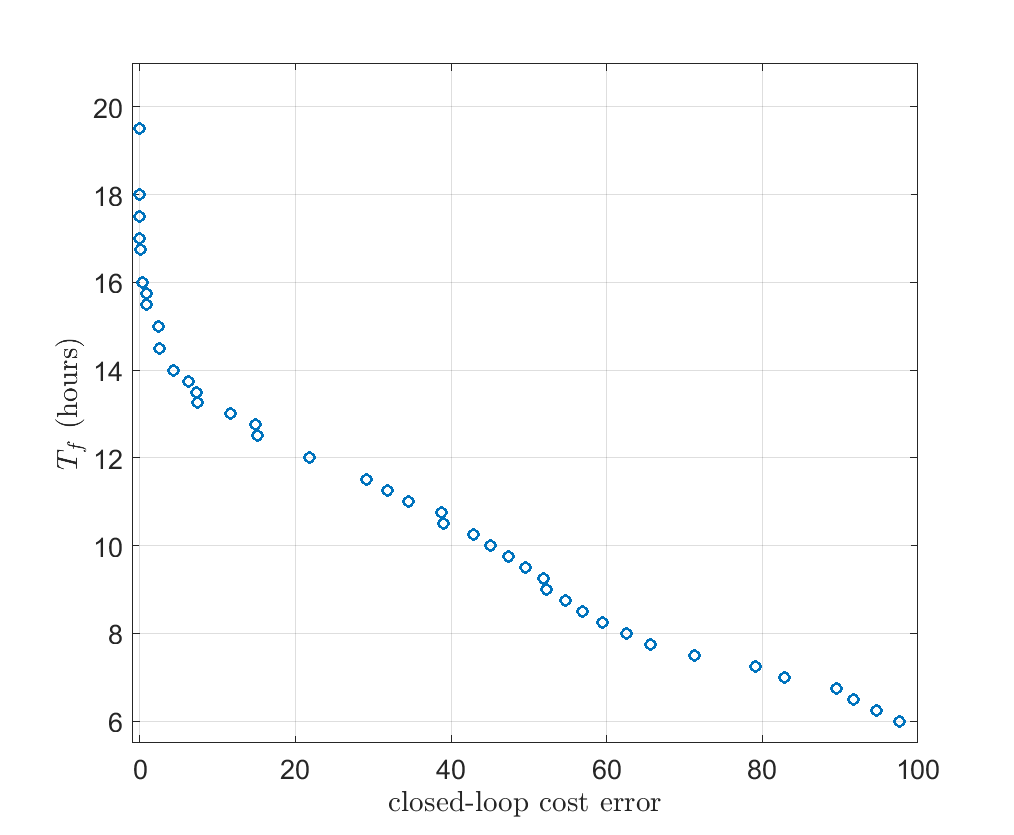}
    \caption{Prediction horizon tuning for $T^{(d)}_k=T^{(s)}=15$min}
    	\label{Fig_pareto2}	
\end{figure}

Figure \ref{Fig_paretopc} shows the Pareto fronts for two different technology sizes.  
It is interesting to observe that the choice of the MPC parameters depends on the technology size. The co-design requires a controller providing good performance for all the possible sizes. 
Conversely, the controller parameter needs to perform well only for the specific size once the system design is completed.
Consequently, the final decision can compromise accuracy with real-time computational efficiency.
The results indicate that a closed-loop cost error of $\pounds \,0.15$ per week can be achieved by  $T^{(d)}=5$ min  $T^{(s)} =15$ min and $t_f=17$ hours.
Instead, the choice of $T^{(d)}=T^{(s)} =15$ min and $t_f=24$ hours gives a closed-loop cost of  $\pounds \, 1.31 $ per week. 
The Pareto front of the cost error performed only considering the prediction horizon as a tuning parameter is illustrated in Figure \ref{Fig_pareto2}. 
The studies highlight the importance of performing the tuning of the controller in an integrated fashion with the building design. 

\section{Conclusions}
\label{sec:Conclusions}
The achievement of net-zero carbon emissions requires decarbonisation of the entire housing stock. We have presented a robust framework to simultaneously optimise the design, the controller and the operation of residential buildings considering external weather conditions and time-varying electricity prices. 

The simultaneous optimization of the design, control and operation of a building considering uncertainty is a computationally challenging optimization problem. The challenges are primarily related to the multi-objective nature of the operation of a building, as well as to long operating timescales and corresponding exogenous data. The framework proposed in this paper mitigates the computational complexity by decoupling multiple objectives and iteratively solving optimization problems of reduced size. Furthermore, to further improve the complexity, we sample the data sets of the exogenous data to include data with the most critical information for the decision process.

Case studies demonstrate the ability of the presented co-design framework to seek trade-offs in an integrated fashion with a temporal resolution spanning from years to minutes. The results show a sensitivity of the optimal solution to data and initial conditions. 
This high sensitivity also demonstrates that the range of price variations is such that the value of the initial energy stored is comparable to the saving achieved.
It also suggests that the combination of the considered technologies is only convenient in highly dynamic and uncertain electricity markets unless other sources of revenue, such as ancillary services, are accounted for as possible additional income. 

In particular, the case study reported a lower cost for the robust co-design framework than for the deterministic approach in all cases, up to a reduction of about 30\% in some cases.
The developed approximations and solution approaches report a computational time reduction at least $10$ times lower compared to the original problem with a deterioration in the performance of only 0.6\%.

Overall, the flexibility of MPC and the optimal design of residential buildings indicate that the presented framework is a good candidate for future work.
The importance sub-sample approach and the MPC tuning algorithm require further analysis and improvements in terms of their accuracy and computational performances.
Note that the proposed framework is generalizable to other contexts and applications, such as power systems, modern transports, robotics, medical devices and manufacturing processes.

\bibliography{bibio}

\end{document}